\documentclass[12pt,preprint]{aastex}

\shorttitle{Strong Blueshifts Observed with {\it Hinode}/EIS}
\shortauthors{Asai et al.}

\begin{document}

\title{Strongly Blueshifted Phenomena Observed with {\it Hinode}/EIS
in the 2006 December 13 Solar Flare}

\author{
Ayumi Asai\altaffilmark{1,2,3}, 
Hirohisa Hara\altaffilmark{2,3}, 
Tetsuya Watanabe\altaffilmark{2,3}, 
Shinsuke Imada\altaffilmark{2}, \\
Taro Sakao\altaffilmark{4},
Noriyuki Narukage\altaffilmark{4},
J. L. Culhane\altaffilmark{5}, and
G. A. Doschek\altaffilmark{6}}

\email{asai@nro.nao.ac.jp}
\altaffiltext{1}{Nobeyama Solar Radio Observatory, National Astronomical
Observatory of Japan, 
Minamimaki, Minamisaku, Nagano, 384-1305, JAPAN}

\altaffiltext{2}{National Astronomical Observatory of Japan, 
Osawa, Mitaka, Tokyo, 181-8588, JAPAN}

\altaffiltext{3}{The Graduate University for Advanced Studies
(SOKENDAI), Hayama, Miura, Kanagawa, 240-0193, JAPAN}

\altaffiltext{4}{Institute of Space and Astronautical Science, Japan
Aerospace Exploration Agency, Yoshinodai, Sagamihara, Kanagawa,
229-8510, JAPAN}

\altaffiltext{5}{Mullard Space Science Laboratory, University College
London, Holmbury St. Mary, Dorking, Surrey, RH5 6NT, U.K.}

\altaffiltext{6}{Naval Research Laboratory, E.O. Hulburt Center for
Space Research, Washington DC 20375-5320, U.S.A.}


\begin{abstract}
We present a detailed examination of strongly blueshifted emission lines
observed with the EUV Imaging Spectrometer on board the {\it Hinode}
satellite.
We found two kinds of blueshifted phenomenon associated with the X3.4
flare that occurred on 2006 December 13.
One was related to a plasmoid ejection seen in soft X-rays.
It was very bright in all the lines used for the observations.
The other was associated with the faint arc-shaped ejection seen in soft
X-rays. 
The soft X-ray ejection is thought to be an MHD fast-mode shock wave.
This is therefore the first spectroscopic observation of an MHD
fast-mode shock wave associated with a flare.
\end{abstract}

\keywords{Sun: corona --- Sun: flares --- Sun: transition region --- 
Sun: UV radiation --- Sun: X-rays, gamma rays}

\section{Introduction}
Solar flares are very spectacular, and they are accompanied by a variety
of plasma motions.
For example, many ejection phenomena, such as filament/prominence
eruptions are seen in the H$\alpha$ line and in the extreme ultraviolet
(EUV), while plasmoid ejections seen in soft X-ray (SXR) have been
observed in association with solar flares.
They have attracted attention, since they could play a key role in
triggering fast reconnection.
In the plasmoid-induced reconnection model, which was suggested by
\citet{Shiba99} and \citet{Shiba01} as an extension of the classical
CSHKP model \citep{Car64,Stur66,Hira74,Kop76}, plasmoids generated in
current sheets near reconnection points are ejected, when strong energy
releases occur.
The ejections of plasmoids trigger further reconnections, since they
make the current sheet thinner.
We often observe that plasmoids/filaments are strongly accelerated
during bursts of nonthermal emissions in hard X-ray (HXR) and in radio
\citep{Kah88,Shiba95,Tsu97,Ohya98,Mori03,Ster04,Ster05}.
Therefore, the correlations between plasmoid ejections and HXR bursts
support a plasmoid-induced reconnection model.
A good review of the correlation between plasmoid ejections and HXR
emission is given in \citet{Asch02}.

Flare-associated waves and the related plasma motions have also been
studied.
Moreton waves \citep{More60,Smi71} are observed to propagate across the
solar disk in H$\alpha$ with speeds of 500 -- 1500~km~s$^{-1}$
(e.g., \cite{Eto02,War04a,War04b}).
They are often associated with type II radio bursts, and thought to be
due to the intersection of a coronal MHD fast-mode shock wave and the
chromosphere \citep{Uchi68,Uchi70,Uchi73}.
X-ray waves discovered with the Soft X-ray Telescope (SXT; \cite{Tsu91})
on board {\it Yohkoh} \citep{Oga91} are the wavelike disturbances
travelling in the solar corona associated with flares
\citep{Kha00,Kha02,Hud03}.
The simultaneous observation of Moreton waves and X-ray waves suggest
that both of these wavelike disturbances are generated with the MHD
fast-mode shock \citep{Kha02,Naru02}.
Furthermore, the Solar X-Ray Imager on board {\it GOES} has also
observed wavelike disturbances in X-rays \citep{War05a}.
Coronal EIT waves observed with the Extreme-Ultraviolet Imaging
Telescope (EIT; \cite{Dela95}) on board the {\it Solar and Heliospheric
Observatory} ({\it SOHO}; \cite{Dom95}) are another coronal disturbance
(e.g., \cite{Thom00}).
However, these coronal disturbances have rarely been observed
spectroscopically in EUV and in SXR.
Therefore, spectroscopic observations of wavelike ejections with high
spatial and spectral resolution are required to clarify the relation
with the MHD fast-mode shock.

These plasma motions can be observed as phenomena accompanied by line
shifts (Doppler shifts) in spectroscopic observations.
The EUV Imaging Spectrometer (EIS; \cite{Cul07}) is one of the three
scientific instruments on board {\it Hinode} \citep{Kosu07}.
EIS uses an off-axis parabolic primary mirror and a toroidal diffraction
grating in a normal incidence optical layout, and has sensitivity for
two wavelength ranges, 170 -- 210 and 250 -- 290~{\AA} (see
\cite{Kore06} and \cite{Lang06}, for more details).
These wavelength ranges are simultaneously observed with two CCDs, which
are called CCD-A (for longer wavelength range) and CCD-B (for shorter
wavelength range), respectively.
The two dimensional EUV images were obtained with the narrow slit in a
raster observation by a pivot rotation of the primary mirror in the
east-west direction.
EIS enables us to study in detail the plasma in the solar corona and
upper transition region with temperature of $8\times10^4$~K to
$2\times10^7$~K.

We found strongly blueshifted phenomena, which were associated with an
intense flare that occurred 2006 December 13 observed with EIS.
In this paper we examined the phenomena, and discuss the relation with
plasmoid ejections and/or coronal waves.
In \S 2 we describe the observational data.
In \S 3 we examine the detailed features of the strongly blueshifted
line emission.
In \S 4 we present a summary and conclusions.

\section{Observations}
An intense solar flare, which was X3.4 on the GOES scale, started at
02:14~UT, 2006 December 13, in NOAA Active Region 10930 (S06$^{\circ}$,
W22$^{\circ}$).
It was the first X-class flare that {\it Hinode} observed (see also
\cite{Kubo07}, \cite{Su07}, \cite{Ima07}, \cite{Isobe07}).
The top panel of Figure~1 shows time profiles of the flare (from 02:10
UT to 03:00 UT, 2006 December 13), in SXR obtained with the GOES 1.0 --
8.0 (top) and 0.5 -- 4.0 {\AA} (bottom) channels.
The Nobeyama Radio Polarimeters (NoRP; \cite{Torii79,Shiba79,Naka85})
also observed this flare.
The bottom panel of Figure~1 shows the time profiles of the total radio
fluxes taken in NoRP~17 GHz (black line) and 35~GHz (gray line)
channels.

EIS performed a wide raster scan from 01:12 to 05:41~UT, and the
following nine lines were chosen to observe the flare:
Fe {\sc x} (184.5 {\AA}), Fe {\sc viii} (185.2 {\AA}), 
Fe {\sc xi} (188.2 {\AA}), Ca {\sc xvii} (192.8 {\AA}),
Fe {\sc xii} (195.1 {\AA}), Fe {\sc xiii} (202.0 {\AA}), 
He {\sc ii} (256.3 {\AA}), Fe {\sc xiv} (274.2 {\AA}), and 
Fe {\sc xv}(284.2 {\AA}).
Information on these lines is summarized in Table~1.
We used a narrow slit of 1$^{\prime\prime}$ width.
For each emission line, we set the observation window on the CCDs with a
height of 256 pixels along the slit and a width of 24 pixels in the
wavelength direction.
These correspond to 256$^{\prime\prime}$ and 0.54~{\AA}, respectively.
We show the raster images in the He~{\sc ii} and Fe~{\sc xv} lines in
the bottom panels of Figure~2.
The field of view (FOV) of CCD-A is displaced southward in the slit
direction by about 18$^{\prime\prime}$, compared with that of CCD-B.
In this work we have corrected for this displacement.
The absolute positions of the FOVs are determined with the EUV images
observed with other instruments as will be described below.

We are focusing on the strongly blueshifted feature observed with EIS.
In the impulsive phase two kinds of feature are observed.
One is located about $120^{\prime\prime}$ south of the disk center,
which is just south of the flare core region.
This is bright in all lines, and drifts southward with a velocity of
about 50~km~s$^{-1}$.
The small box drawn with black dotted lines in the bottom right panel of
Figure~2 shows the position of this feature.
The other blueshifted feature appears farther from the flare core site
and is located $200^{\prime\prime}$ south of the disk center.
This very faint feature is seen only in high-temperature lines, and
moves southward with high velocity.
We call these features ``BS1'' and ``BS2'', and they are discussed in
more detail in the next section.

The imaging observation of this flare in EUV was performed with the
Transition Region and Coronal Explorer (TRACE; \cite{Hand99,Sch99})
although the impulsive phase was unfortunately missed.
We show the 195~{\AA} images of the flare taken by TRACE at 02:05:25~UT
(left) and 02:47:20~UT (right) in the top panels of Figure~2.
We can see a dark filament lying horizontally (i.e., in the east-west
direction) in the western part of the active region.
It can  also be seen in the EIS raster images until 02:20~UT.
The dark filament disappeared after the flare started.
SOHO/EIT also observed the eruption of the filament in the 195~{\AA}
images, travelling in the southwest direction.
We co-aligned the EUV images obtained with EIS, TRACE, and EIT, by using
the 195~{\AA} data.
In particular, the EIS raster image was fitted with the TRACE preflare
(02:05:25~UT) image, by using the common features seen from 02:00 to
02:10~UT.
The accuracy of the co-alignment between the EIS and TRACE data for the
time range is about 1$^{\prime\prime}$, which is comparable to the pixel
sizes of the data.

The Solar Optical Telescope (SOT; \cite{Tsu07}) and X-Ray Telescope
(XRT; \cite{Gol07}) on board {\it Hinode} also observed this flare.
The Ca~{\sc ii} (H-line) images obtained every 2 minutes with SOT
clearly show the two-ribbon structure (see the top panels of Figure~4).
XRT obtained the SXR images of this flare every 1 minute with the
Thin-Be filter.
In the SXR images we can see some ejections of bright plasmoids (see
the bottom panel of Figure~4).
We can also see a faint arc-shaped ejection (see the bottom panels of
Figure~5).
We will discuss these in the following section.
We co-aligned the SOT and XRT data with the TRACE data. 
For the co-alignment, we used common features, such as flare kernels.
The accuracy of the co-alignment is also about 1$^{\prime\prime}$, and
therefore, we expect that the EIS data are co-aligned with the SOT and
XRT data with an accuracy of about 2$^{\prime\prime}$.

\section{Blueshifted Features}
\subsection{The Northern Feature -- BS1}
This strong blueshifted feature associated with BS1 appeared from
02:23:46~UT to 02:26:24~UT in the EIS FOV, which corresponds to the
first impulsive radio burst observed with NoRP (see Fig.~1).

The top panels of Figure~3 show the clipped spectra of BS1 in the
He~{\sc ii} (left panel) and Fe~{\sc xv} (right panel) lines.
The vertical axis is the extent along the slit (the solar north is up),
and the size is 224$^{\prime\prime}$ ($\sim 1.6\times10^{5}$~km).
The horizontal axis shows the wavelength direction (blue is to the
left), and the width is 24 pixels in the CCDs of EIS, which corresponds
to about 0.54~{\AA} in the wavelength range.
The spectra are taken at (A) 02:24:49~UT, (B) 02:25:21~UT, (C)
02:25:52~UT, (D) 02:26:24~UT, and (E) 02:26:56~UT, respectively.
The flare core site is seen as very bright horizontal bands in the
spectra, and it is saturated in the He~{\sc ii} line.
BS1 is located just south of the flare core region, and it drifts in the
slit direction (southward) with a velocity of about 50~km~s$^{-1}$.
The flare cores are seen as blob-like bright features in all the lines
used for the raster.

The bottom panels of Figure~3 show the spectra of the blueshifted
regions in the He~{\sc ii} (left) and Fe~{\sc xv} (right) windows.
The histograms shown with the solid lines are the observed spectra
integrated over the blueshifted region sandwiched between the two white
horizontal lines in the top panels for each emission line.
The spectra are normalized with their maximum intensities.
We fitted the spectra with Gaussian functions.
For each window, the spectrum is divided into a main component (dotted
line) and a blueshifted component (solid line).
The black thin and thick arrows point to the peaks of the main and
blueshifted components, respectively.
The Doppler velocity is determined by the displacement between the
blueshifted and the main components.
Although the main components themselves show displacements compared to
the profiles of a quiet region (e.g., the bottom part of the FOVs) of
less than 5~km~s$^{-1}$, they are small compared with the blueshifted
phenomena, and therefore, we do not take them into the following
considerations.
The He~{\sc ii} and Fe~{\sc xv} lines recorded Doppler velocities of
about 280 and 240~km~s$^{-1}$, respectively.
For BS1, the blueshifted components are brighter than the main
components.

We compared the features with the SOT data to investigate the relation
between BS1 and the flare kernels.
We often observe upflow motion, associated with chromospheric
evaporation \citep{Neu68,Hira74,Anti78,Anto82,Can85}, at the footpoints
of flare loops.
They result from sudden pressure enhancement due to bombardment by
nonthermal particles and/or conduction from the coronal flare kernels.
\citet{Milli06} reported that the upflows of about 110~km~s$^{-1}$ in
the Fe~{\sc xix} line (log$T$ $\approx$ 6.9) were observed with the {\it
Reuven Ramaty High-Energy Solar Spectroscopic Imager} ({\it RHESSI}:
\cite{Lin02})at the flare kernels.

The top panels of Figure~4 show the Ca~{\sc ii} (H-line) images.
We can see a two-ribbon structure that shows an inverse S-shape in the
east-west direction.
The crosses in the panels ($\times$) point to the positions of BS1
observed with EIS.
The X-coordinate corresponds to the slit position at that time, and the
Y-coordinate is the front position (i.e., the southern end) of BS1 seen
in the Fe~{\sc xv} line.
We cannot see any correspondences between BS1 and the bright features
seen in the SOT images, such as the flare ribbons.
Therefore, BS1 is not related to the evaporation flow.

We compared the EIS data with the SXR images obtained with XRT.
The bottom panels of Figure~4 show these images.
As we mentioned, we can see a plasmoid ejection as marked with arrows.
It lies horizontally, and is ejected from the flare core site in the
southwest direction.
The crosses in the panels again indicate the front positions of BS1
determined with the Fe~{\sc xv} window.
We can see a remarkable correspondence in all the frames between
the positions of BS1 and those of the plasmoid ejection.
The ejected plasmoid travels in the XRT images with a velocity of about
90~km~s$^{-1}$, which is roughly consistent with the drift speed of BS1.
Therefore, we conclude that BS1 originates from the plasmoid ejection.

As we mentioned above, the timing of BS1 just corresponds to the first
impulsive radio burst observed with NoRP, which supports the
plasmoid-induced reconnection model.
BS1 showed Doppler and drift velocities of about 240 -- 280 and 50 --
90~km~s$^{-1}$, respectively, which indicates that the combined velocity
was about 250 -- 300~km~s$^{-1}$.
This value is consistent with the velocities that have been previously
reported (e.g., \citet{Shiba95} examined eight X-ray plasmoid ejections
observed with {\it Yohkoh}, and found that the apparent velocity was 50
-- 400~km~s$^{-1}$).

\subsection{The Southern Feature -- BS2}
The top panels of Figure~5 show time sequences of the spectra in the
Fe~{\sc xv} and Ca~{\sc xvii} windows.
Solar north is up and shorter wavelength (i.e., blueward) is to the left.
The size of each spectrum is 224$^{\prime\prime}$ along the slit
(vertical), and 24 pixels in the EIS CCDs (horizontal).
The displacement between the CCD-A and CCD-B was corrected.
The times of the spectra are taken at (a) 02:22:11~UT, (b) 02:22:43~UT,
(c) 02:23:14~UT, (d) 02:23:46~UT, and (e) 02:24:18~UT, respectively.

The flare core site is seen as two very bright bands in these spectra.
BS2 appeared from 02:22:11~UT to 02:24:18~UT, and went out of the EIS
FOV just before the first impulsive radio burst was observed with NoRP
(see Figure~1).
They started to appear at $200^{\prime\prime}$ south of the disk center,
which is about $100^{\prime\prime}$ south of the flare core region.
They travel along the slit, i.e., southward, with a velocity of about
450~km s$^{-1}$.
Furthermore, as seen in the top panels of Figure~5, these blueshifted
features broaden widely in the blue wings of the lines. 
These features were observed only in the Fe~{\sc xv} and Ca~{\sc xvii}
windows, which are the two hottest lines contained in the raster.
Therefore, the plasma generating BS2 must be heated more than about
2~MK.

The middle panels of Figure~5 show the spectra of the blueshifted
regions in the Fe~{\sc xv} and Ca~{\sc xvii} windows.
The histograms shown with solid lines are the observed spectra
integrated over the region sandwiched between the two white lines in the
top panels for each line.
The spectra are normalized with the maximum intensities.
We fitted the spectra and divided them into several components.
We assumed that all the components follow Gaussian functions.

For the Fe~{\sc xv} window, the spectrum is divided into Fe~{\sc xv}
main component (dotted line) and the blueshifted component (solid line).
The black thin and thick arrows point to the peaks of the main and
blueshifted components, respectively.
The Ca~{\sc xvii} window (the middle right panel) contains Fe~{\sc xi}
and O~{\sc v} lines as well as the target Ca~{\sc xvii} line.
We can see Fe~{\sc xi} and O~{\sc v} lines discretely in the quiet
region, while we can see the blended emission with the Ca~{\sc xvii}
lines in the BS2 region.
Therefore, we have to remove those components, before we discuss the
blueshift component of the Ca~{\sc xvii} line.
First, we do not take the O~{\sc v} component into consideration, since
we cannot see the component at all for the blueshifted region.
The line is sensitive to the low temperature plasma (log$T$ is 5.4), and 
the absence is consistent with the fact that BS2 is hot.
Secondly, we estimate the Fe~{\sc xi} component in the Ca~{\sc xvii}
window.
Since the Fe~{\sc xi} emission line in the Ca~{\sc xvii} window
(192.8~{\AA}) and the left peak (i.e., the peak with the shorter
wavelength) of the Fe~{\sc xi} doublet (188.2~{\AA}) are a density
insensitive line pair, we can estimate the intensity and the line shift
compared to the quiet region from the profile of the Fe~{\sc xi} lines
in the Fe~{\sc xi} window.
In the middle right panel of Figure~5 we show them with the gray dashed
line.
The Fe~{\sc xi} component also do not show blueshift at all.
The gray arrows point to the peak positions of the Fe~{\sc xi} in the
Ca~{\sc xvii} window.
Then, after the subtraction of the Fe~{\sc xi} component, we divided the
spectrum into Ca~{\sc xvii} main component (dotted line) and the
blueshifted component (solid line).
The black thin and thick arrows again point to the peaks of the Ca~{\sc
xvii} main and the blueshifted components, respectively.

The peaks of the BS2 components shifted 0.089~{\AA} for the Fe~{\sc xv}
line, and 0.099~{\AA} for the Ca~{\sc xvii} line, which correspond to
Doppler velocities of about 90 and 150~km~s$^{-1}$, respectively.
The blueshifted components of both lines broaden widely, and have wide
FWHM of about 0.24~{\AA} for the Fe~{\sc xv} line and about 0.21~{\AA}
for the Ca~{\sc xvii} line.
The very wide FWHM of the blueshifted components indicates that BS2 has
a large velocity field along the line of sight.
These features in the spectra are very different from those of BS1.

We compared BS2 with the coronal features seen in the XRT images.
We expect plasma motion in the XRT images at the same position as BS2,
since XRT is sensitive to high temperature plasma observed with the EIS
Fe~{\sc xv} and Ca~{\sc xvii} spectra.
The bottom three panels of Figure~5 show the XRT Thin-Be filter images
of the flare.
The arrows point to the front of the XRT arc-shaped ejection, which is
travelling in the southeast direction with a velocity of about
650~km~s$^{-1}$.
This ejection was faint compared with the other ejections seen in the
XRT and TRACE images.
The shape and the travelling direction are also different from the others
which showed the rod-shaped structure and traveled in the southwest
direction, similar to that associated with BS1.
These features of the ejection resemble the X-ray waves discovered with
{\it Yohkoh}/SXT, and this disturbance is possibly generated by the MHD
fast-mode shock.
We roughly estimated the Alfv\'{e}nic Mach number of BS2, using the same
method as \citet{Naru02}, and found it to be about 1.4.
This result indicates that the XRT wavelike phenomena and BS2 might be a
weak MHD fast-mode shock.
The detailed analysis and theoretical treatment of the shock is beyond
the scope of this paper.
The observational confirmation of the density and temperature jumps
caused by the passage of the shock waves will be discussed in a future
paper.

The crosses in the panels show the position of BS2.
The horizontal positions (X-coordinate) correspond to the EIS slit
position.
We can clearly see the correspondence between BS2 and the ejection as
part of a coronal shock wave, and therefore, we conclude that BS2 is
associated with the wavelike phenomenon.
Furthermore, a Halo-type coronal mass ejection (CME) associated with the
flare was observed with the Large Angle Spectrometric Coronagraph
(LASCO) on board the {\it Solar and Heliospheric Observatory} ({\it
SOHO}: see the {\it SOHO}/LASCO CME online catalog\footnote{See
http://cdaw.gsfc.nasa.gov/CME\_list/.}; \cite{Yashi04}).
The EIT wave travels in almost all direction from the flare site, and
the clearest disturbance is seen travelling with a speed of about 570
$\pm$ 150 km s$^{-1}$in the southeast direction, which is almost the
same as that of the XRT wave.
Figure~6 shows the running differences of the EIT images.
The 02:24:33~UT image (bottom left panel) suffered from scattered light
in the telescope, and therefore, the wave-front is unclear.
In the XRT image (top left panel) the front of the XRT wave is indicated
by the solid line, and the wave is estimated to reach the dashed line
position at 02:36~UT.
In the 02:36:01~UT image (bottom right panel) the expected position of
the XRT wave is also shown by the dashed line, and we can see that the
front of the EIT wave is located slightly inside of the XRT wave
position (dashed line).

BS2 recorded a Doppler velocity of about 100~km~s$^{-1}$ while the drift
velocity along the slit is about 450~km~s$^{-1}$.
This implies that BS2 was travelling with a velocity of about
460~km~s$^{-1}$ and with an elevation angle of 13 degree from the plane
of sky, although the Doppler velocity (100~km~s$^{-1}$) was derived only
from two points of measurement.
The velocity of 460~km~s$^{-1}$ is slower than the typical velocity of
Moreton waves and/or X-ray waves.
This is because the slit direction (north-south) is different from the
direction of the arc-shaped ejection.

\section{Summary and Conclusions}
We found two kinds of strongly blueshifted feature that were observed
with {\it Hinode}/EIS to be associated with the 2006 December 13 flare.

BS1 was bright in all observed lines used for the raster, and drifted
southward, that is along the slit, at a velocity of about
50~km~s$^{-1}$.
The Doppler velocity is about 250~km~s$^{-1}$.
They are associated with the plasmoid ejection seen in XRT, while there
are no corresponding flare ribbons in the Ca~{\sc ii} (H-line) images
obtained with SOT.
Therefore, we concluded that BS1 is the ejected plasma, and is not an
evaporation flow.
Moreover, BS1 appears at the time of the first radio burst, which
supports the plasmoid-induced reconnection model that shows the
correlation between acceleration of plasmoids/filaments and bursts of
nonthermal emissions.

BS2, on the other hand, was very faint, and showed spectra that
broaden in the wavelength space.
The center of the blueshifted component recorded a Doppler velocity of
about 100~km~s$^{-1}$, the drift velocity along the slit is about
450~km~s$^{-1}$.
These components are observed only in the hottest lines of the raster
observation (Fe~{\sc xv} and Ca~{\sc xvii}), and therefore, the plasmas
must be heated to more than 2~MK.
The BS2 region corresponds to the propagation of the coronal wavelike
ejection seen in XRT images.
The ejection is thought to be a MHD fast-mode shock wave, and it is the
first successful spectroscopic observation of such a shock wave
associated with a flare.

\acknowledgments

We first acknowledge an anonymous referee for her/his useful comments 
and suggestions.
{\it Hinode} is a Japanese mission developed and launched by ISAS/JAXA,
collaborating with NAOJ as domestic partner and NASA and STFC (U.K.) as
international partners.
Scientific operation of the {\it Hinode} mission is conducted by the
{\it Hinode} science team organized at ISAS/JAXA.
This team mainly consists of scientists from institutes in the partner
countries.
Support for the post-launch operation is provided by JAXA and NAOJ
(Japan), STFC (U.K.), NASA, ESA, and NSC (Norway).
This work was partly carried out at the NAOJ Hinode Science Center,
which is supported by the Grant-in-Aid for Creative Scientific Research
''The Basic Study of Space Weather Prediction'' from MEXT, Japan
(17GS0208, Head Investigator: K. Shibata), generous donations from Sun
Microsystems, and NAOJ internal funding.
JLC acknowledges the award of a Leverhulme Emeritus Fellowship.


\clearpage

\begin{table}
\begin{center}
Table 1 \\
Line list used of the raster observation\\
\begin{tabular}{lllc} \tableline\tableline
Target Ion & Center wavelength ({\AA}) & CCD & log$T$ \\ 
\tableline
Fe {\sc x}   & 184.5 & CCD-B & 6.0 \\
Fe {\sc viii}& 185.2 & CCD-B & 5.6 \\
Fe {\sc xi}  & 188.2 & CCD-B & 6.1 \\
Ca {\sc xvii}& 192.8 & CCD-B & 6.8 \\
Fe {\sc xii} & 195.1 & CCD-B & 6.2 \\
Fe {\sc xiii}& 202.0 & CCD-B & 6.2 \\
He {\sc ii}  & 256.3 & CCD-A & 4.9 \\
Fe {\sc xiv} & 274.2 & CCD-A & 6.2 \\
Fe {\sc xv}  & 284.2 & CCD-A & 6.4 \\
\tableline
\end{tabular}
\end{center}
\end{table}

\begin{figure}
\epsscale{0.5}
\plotone{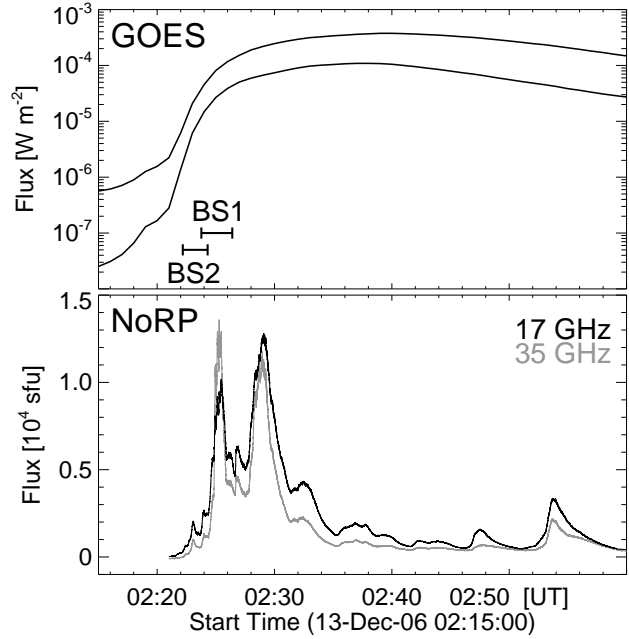}
  \caption{
Overview of the 2006 December 13 flare.
Temporal variations of the GOES SXR flux in the 1.0 -- 8.0~{\AA} and 0.5
-- 4.0 channels ({\it top}) and radio time profiles (17 and 35~GHz)
obtained with NoRP ({\it bottom}).
The time ranges of BS1 and BS2 seen in the EIS field of view are marked
with BS1 and BS2, respectively.}
\label{fig:2}
\end{figure}

\begin{figure}
\epsscale{.75}
\plotone{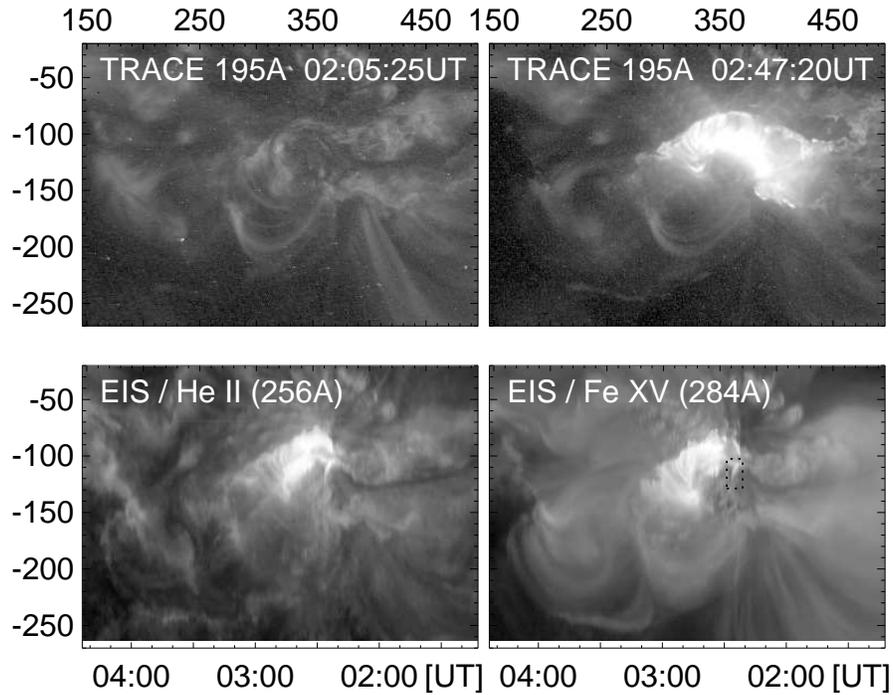}
  \caption{
EUV (195 {\AA}) images taken with TRACE ({\it top}) and raster images of
He {\sc ii} (256~{\AA}: {\it bottom left}) and in Fe {\sc xv}
(284~{\AA}: {\it bottom right}) lines.
Solar north is up, and west is to the right.
The horizontal and vertical axes give the distance from the disk center
in arcseconds as shown in the top panels.
For the EIS raster images, the horizontal axis below the panels shows
the times.
The small box with the dotted line in the bottom right panel points to
the BS1 region.}
\label{fig:2}
\end{figure}

\begin{figure}
\epsscale{.75}
\plotone{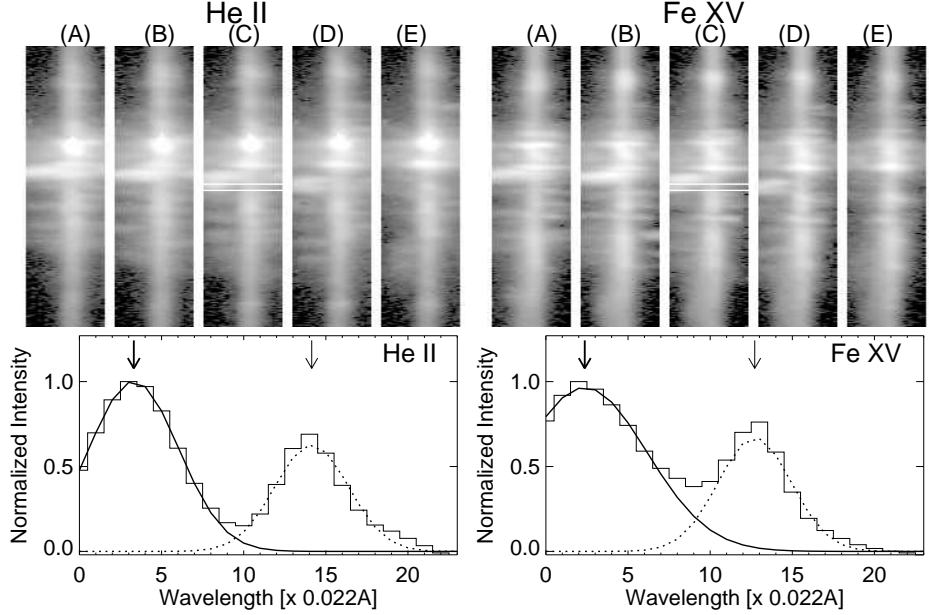}
  \caption{
Northern blueshift (BS1).
{\it Top}: Time sequenced spectra of He~{\sc ii} (left) and Fe~{\sc xv}
(right) windows observed with {\it Hinode}/EIS.
The time of each panel is (A) 02:24:49~UT, (B) 02:25:21~UT, (C)
02:25:52~UT, (D) 02:26:24~UT, and (E) 02:26:56~UT.
Solar north is up and blue is to the left.
The size of each window is 224$^{\prime\prime}$ along the EIS slit
(vertical) and 24 pixels in the CCDs of EIS, which corresponds to
0.54~{\AA} in the wavelength scale (horizontal).
{\it Bottom}: Normalized spectra at BS1 in He~{\sc  ii} (left) and
Fe~{\sc xv} (right) windows.
The solid histograms show the spectra averaged over the region
sandwiched between the two horizontal white lines in the top panels.
The dotted and solid lines are the fitting results that represent the
main and the blueshift components of the line.
The peaks of each line are shown with thin and thick arrows.}
\label{fig:3}
\end{figure}

\begin{figure}
\epsscale{.50}
\plotone{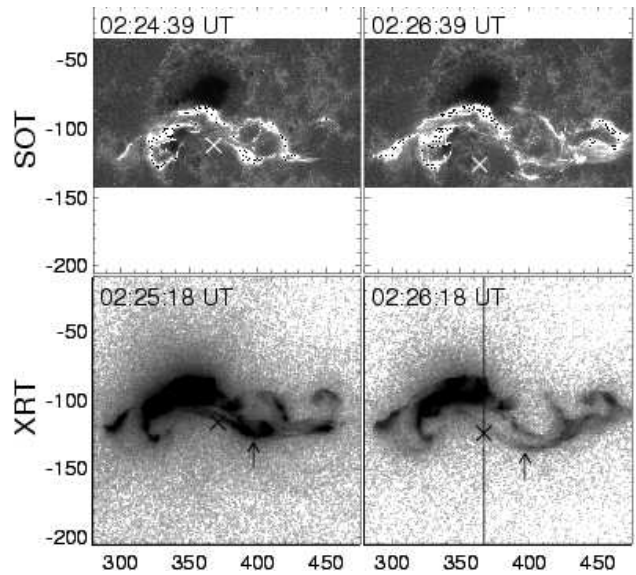}
  \caption{
{\it Top}: Ca~{\sc ii} (H-line) images taken with {\it Hinode}/SOT.
{\it Bottom}: Soft X-ray images obtained with the {\it Hinode}/XRT
(negative images).
Solar north is up, and west is to the right.
The horizontal and vertical axes give the distance from the disk center
in arcseconds as shown in the top and the left.
The cross signs ($\times$) point to the positions of BS1.
The arrows in the bottom right panel follows the plasmoid ejection seen
in XRT.
The vertical line in the bottom right panel shows the position of the
EIS slit.
The slit extends from $y=-233$ to $-9$ arcseconds.}
\label{fig:4}
\end{figure}

\begin{figure}
\epsscale{.75}
\plotone{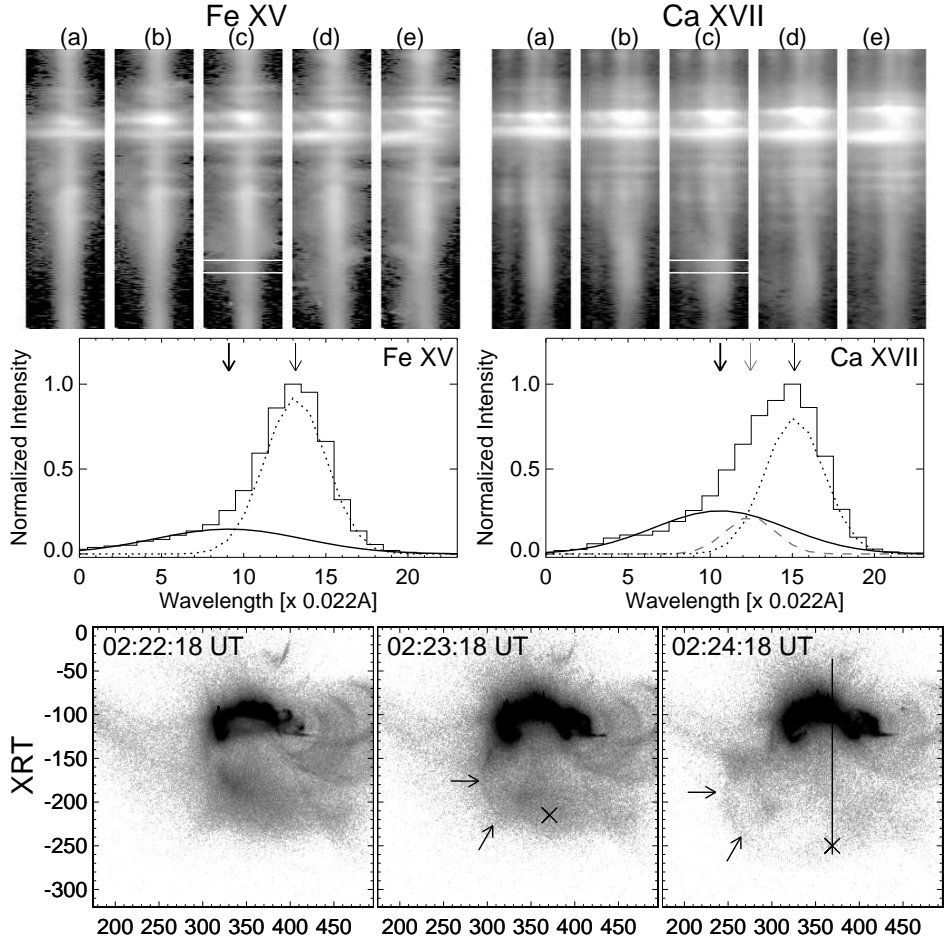}
  \caption{
Southern blueshift (BS2).
{\it Top}: Time sequenced spectra of Fe~{\sc xv} ({\it left}) and
 Ca~{\sc xvii} ({\it right}) windows observed with {\it Hinode}/EIS.
The time of each panel is (a) 02:22:11~UT, (b) 02:22:43~UT, 
(c) 02:23:14~UT, (d) 02:23:46~UT, and (e) 02:24:18~UT.
Solar north is up and blue is to the left.
The size of each window is 224$^{\prime\prime}$ along the EIS slit
(vertical) and 24 pixels in the CCDs of EIS, which corresponds to 0.54
{\AA} in the wavelength scale (horizontal).
The displacement between the CCD-A and CCD-B was corrected.
{\it Middle}: Normalized spectra at BS2 in Fe~{\sc  xv} (left) and Ca
{\sc~xvii} (right) windows.
The solid histograms show the spectra averaged over the region
sandwiched between the two horizontal white lines in the top panels.
The dotted and solid black lines are the fitting results that represent
the main and the blueshift components of the line.
The peaks of each line are shown with thin and thick arrows.
The gray dashed line in the right panel show Fe~{\sc xi} component, and
the peaks are indicated by gray thin arrows.
{\it Bottom}: Soft X-ray negative images taken with the {\it
Hinode}/XRT.
The arrows show the front of the wavelike ejection.
Crosses ($\times$) represent BS2 determined by the Fe~{\sc xv} line.
The vertical line in the bottom right panel shows the position of the
EIS slit.}
\label{fig:5}
\end{figure}

\begin{figure}
\epsscale{.75}
\plotone{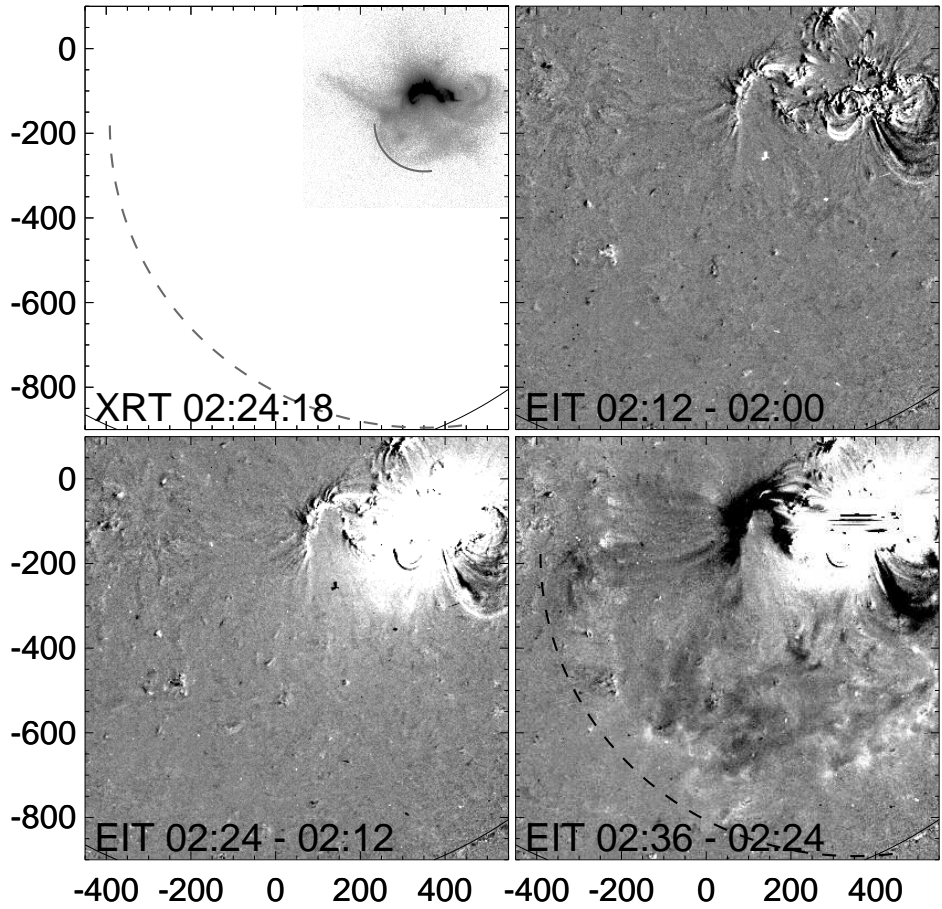}
  \caption{
An XRT image (top left panel) and running differences of EIT images (the
others).
The top right panel shows the image at 02:12~UT with that of 02:00~UT
subtracted.
Differencing is continued in the remaining panels.
The white oval region in the bottom left panel is mainly due to
scattered light in the telescope.
In the top left panel the XRT arc-shaped ejection is indicated by the
solid line, and it is estimated to reach at the dashed line at
02:36~UT.
The estimated position is indicated in the bottom right panel.}
\label{fig:6}
\end{figure}

\end{document}